\documentclass[preprint,aps]{revtex4}
\usepackage{graphics}
\begin{document}
\draft

\title{ 	Complementarity between Gauge-Boson Compositeness
\\ 		and Asymptotic Freedom --- with scalar matter}
\author{		Keiichi Akama}
\address{	Department of Physics, Saitama Medical College,
		Kawakado, Moroyama, Saitama, 350-0496, Japan}
\author{		Takashi Hattori }
\address{	Department of Physics, Kanagawa Dental College,
		Inaoka-cho, Yokosuka, Kanagawa, 238-8580, Japan}

\begin{abstract}
We derive and solve the compositeness condition 
	for the SU($N_c$) gauge boson coupling with $N_s$ scalar fields
	at the next-to-leading order in $1/N_s$ 
	and the leading order in $\ln\Lambda ^2$
	($\Lambda $ is the compositeness scale).
It turns out that 
	the argument of gauge-boson compositeness (with a large $\Lambda $)
	is successful only when $N_s/N_c>22$,
	in which the asymptotic freedom fails,
	as is in the previously investigated case with fermionic matter.
\end{abstract}

 % 11.15.-q Gauge field theories
 % 11.15.Pg Expansions for large numbers of components (
 %          e.g., 1/N sub c expansions)
 % 12.60.Rc Composite models
 % 11.10.Jj Asymptotic problems and properties

\maketitle

The gauge theories have been widely applied not only in high energy physics,
	but also in various branches of physics,
	and the gauge bosons often appear as composites 
	of matter fields \cite{cg} -- \cite{geo}.
For example, some people considered the composite models of weak bosons
	and other gauge bosons \cite{cgw},
	some developed theory of induced gravity 
	with composite graviton \cite{indgra},
	some interpreted the vector mesons (which is known to be composite) 
	as gauge bosons with hidden local symmetry \cite{hidden},
	and others extensively studied dynamically induced `connections' 
	(i.\ e.\ gauge fields) in models of spacetime \cite{emb,qm}, and 
	in molecular, nuclear, and solid-state systems \cite{geo}.
A gauge boson interacting with matter
	can be interpreted as composite
	under the compositeness condition $Z_3=0$ \cite{cc},
	where $Z_3$ is the wave-function renormalization constant
	of the gauge boson.
The compositeness condition gives relations between the effective
	coupling constants and the compositeness scale $\Lambda $.
In our previous papers \cite{ccA} -- \cite{cgb2}, 
	we derived and solved the compositeness condition 
	at the next-to-leading order in $1/N$ in various models,
	and investigated its implications,
	where $N$ is number of the matter field species.
In particular in \cite{cgb2}, 
	considering compositeness of the SU($N_c$) gauge boson 
	coupling with $N_f$ fermion fields,
	we found a ``complementarity" 
	between the gauge-boson compositeness 
	and asymptotic freedom \cite{af} of the gauge theory.
Namely, the gauge-boson can be considered as a composite due to $Z_3=0$
	only when $N_f/N_c>11/2$, in which the asymptotic freedom fails.
Thus it is urgent for us to investigate if such a complementarity is
	accidental or really universal, 
	by, for example, examining other cases of theories.
In this paper, 
	we perform a similar investigation 
	for the SU($N_c$) gauge boson coupling with $N_s$ scalar fields
	instead of the fermion fields.
We find that the complementarity again holds,
	i.e. that the gauge-boson is considered as a composite 
	only when $N_s/N_c>22$, in which the asymptotic freedom fails.

We consider the SU($N_c$) gauge theory 
	for the gauge boson $G_\mu ^a(a=1,\cdots ,N_c^2-1)$ coupling
	with $N_s$ complex scalar fields $\phi _j(j=1,\cdots ,N_s)$, 
	each of which belongs to the fundamental representation of SU($N_c$): 
\begin{eqnarray} 
	{\cal L}&=& 
	-{1 \over 4} \left(  G_{\mu \nu }^a\right)  ^2  
	+ \sum _j\left[ |{\cal D}_\mu \phi _j|^2 
        - m^2|\phi _j|^2 -\lambda (|\phi _j|^2)^2\right] 
\cr &&
	-{1 \over 2\alpha } \left(  \partial ^\mu  G_\mu ^a\right)  ^2 
        +\partial ^\mu  \eta ^{a\dagger } 
         \left(  \partial _\mu  \eta ^a - g f^{abc} \eta ^b G_\mu ^c\right)  
\label{L}
\end{eqnarray} 
with $G_{\mu \nu }^a = \partial _\mu G_\nu ^a - \partial _\nu  G_\mu ^a
                + g f^{abc} G_\mu ^b G_\nu ^c $ and
	${\cal D}_\mu  \phi _j = 
	\partial _\mu  \phi _j - i g T^a G_\mu ^a \phi _j$,
where	$g$ and $\lambda $ are the coupling constants, 
	$m$ is the mass of $\phi _j$,
	$f^{abc}$ is the structure constant of SU($N_c$),
	$T^a$ is the representation matrix for the basis of 
	the associated Lie algebra su($N_c$),
	$\alpha $ is the gauge fixing parameter, and 
	$\eta ^a(a=1,\cdots ,N_c^2-1)$ is the Fadeev-Popov ghost.
In order to absorb the ultraviolet divergences 
	arising from quantum fluctuations,
	we renormalize the fields, mass, and coupling constants as 
\begin{eqnarray} 
	{\cal L}&=& 
	-{1 \over 4} Z_3 \left(  G_{{\rm r}\mu \nu }^a\right)  ^2  
	+ \sum _j\left[ Z_2|{\cal D}_\mu  \phi _{{\rm r}j}|^2 
        - Z_m   m_{\rm r}^2 |\phi _{{\rm r}j}|^2 
	- Z_\lambda  \lambda _{\rm r}(|\phi _{{\rm r}j}|^2)^2\right] 
\cr &&
	-{Z_3 \over 2Z_\alpha  \alpha _{\rm r}} 
	\left(  \partial ^\mu  G_{{\rm r}\mu }^a\right)  ^2 
        + Z_\eta  \partial ^\mu  \eta _{\rm r}^{a\dagger } 
         \left(  \partial _\mu  \eta _{\rm r}^a 
	 - {Z_1\over Z_2} g_{\rm r} f^{abc} \eta _{\rm r}^b G_{{\rm r}\mu }^c
	\right)  
\label{Lr}
\end{eqnarray} 
with	$G_{{\rm r}\mu \nu }^a 
	= \partial _\mu G_{{\rm r}\nu }^a - \partial _\nu  G_{{\rm r}\mu }^a
	+ (Z_1/Z_2)g_{\rm r} f^{abc} G_{{\rm r}\mu }^b G_{{\rm r}\nu }^c $
and	${\cal D}_\mu  \phi _{{\rm r}j} = \partial _\mu  \phi _{{\rm r}j} 
	- i (Z_1/Z_2) g_{\rm r} T^a G_{{\rm r}\mu }^a \phi _{{\rm r}j}$,
where	the quantities with the index ``r" are the renormalized ones,
	and $Z_1$, $Z_2$, $Z_3$, $Z_\eta $, $Z_m$, $Z_\lambda $, and $Z_\alpha $
	are the renormalization constants.
Since the degree of freedom of the $Z_\alpha $ can be absorbed by 
	that of the non-physical parameter $\alpha _{\rm r}$,
	we simply fix it as $Z_\alpha =Z_3$.

If we impose the compositeness condition $Z_3=0$,
	the kinetic term of $G_{{\rm r}\mu }^a$ disappear, 
	and becomes a non-dynamical auxiliary field, 
	whose Euler equation reduces to a constraint.
The constraint, however, causes a self-interaction of $\phi _{\rm r}$, 
	which gives rise to a composite gauge field.
The kinetic term is reproduced 
	through the quantum effects of the matter fields.
This can be treated in the most rigorous way 
	by following the usual renormalization procedure with (\ref{Lr}),
	and by imposing the compositeness condition $Z_3=0$ after that.
As was argued in the case of the model with fermionic matters \cite{cgb2},
	the compositeness condition itself spoils the ordinary 
	perturbation expansion in the coupling constant $g_{\rm r}$,
	and the appropriate expansion parameter is 
	the inverse of the number of the matter-field species, $1/N_s$.
Here we calculate $Z_3$ based on the Lagrangian (\ref{Lr})
	in the leading and the next-to-leading orders in $1/N_s$,
	solve the condition $Z_3=0$ for the coupling constant $g_{\rm r}$,
	and consider its implications with a particular attention 
	on the complementarity 
	between gauge-boson compositeness and asymptotic freedom.
%	on the compositeness-asymptotic-freedom complementarity
%	which has recently been established 
%	in a model with fermionic matters.

\begin{figure}
\includegraphics{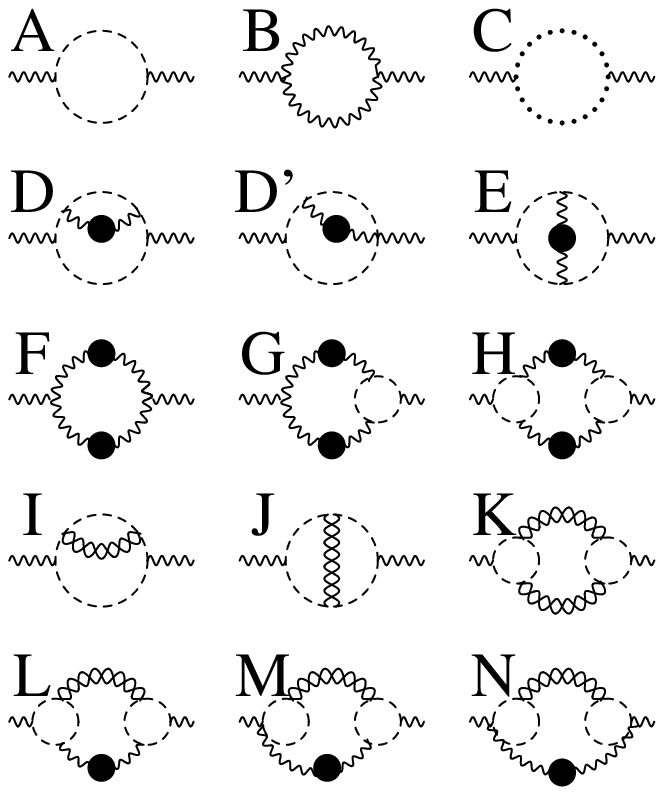}
\caption{The gauge-boson self-energy parts at the leading order (A)
and at the next-to-leading order (B--N) in $1/N_s$.
The dashed, wavy, and dotted lines 
	indicate the elementary scalar, gauge-boson, and Fadeev-Popov ghost 
	propagators, respectively.
The wavy lines with a blob and the chains of small circles are defined in Fig.\ 2. }
\label{f1}
\end{figure}

\begin{figure}
\includegraphics{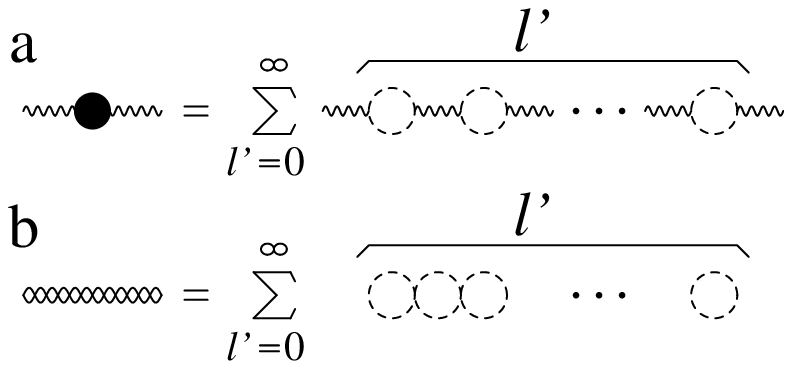}
\caption{ 
The wavy line with a blob 
	stands for the gauge boson propagator 
	with a number of scalar-loops inserted.
The chain of small circles stands for the chain diagram with scalar-loops. 
}
\label{f2}
\end{figure}

\begin{figure}
\includegraphics{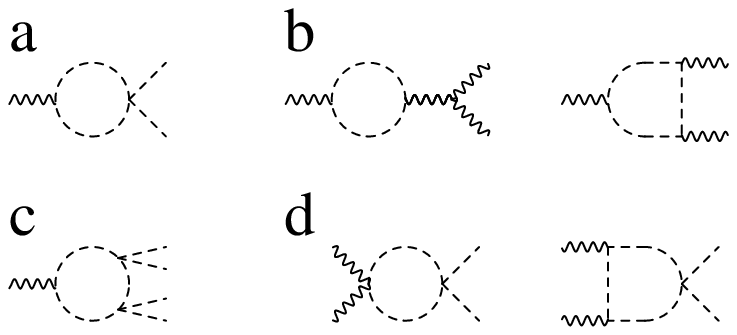}
\caption{ 
The diagram a vanishes, the diagrams in b cancel each other,
the diagram c does not diverge, and the diagrams in d cancel each other.
}
\label{f3}
\end{figure}

In $1/N_s$ expansion, $\lambda$ should be $\le O(1/N_s)$,
	because otherwise diagrams with the more loops become the larger.
We assign $\lambda\sim O(1/N_s)$.
We first choose the renormalization constant $Z_m$
	so as to cancel out the constant (momentum-independent) contributions from 
	the scalar self-mass parts order by order.
%	at the leading and the next-to-leading order in $1/N_s$.
Then the renormalization constant $Z_3$ should be chosen so as to cancel out 
	all the divergences in the diagram A in Fig.\ \ref{f1}
	at the leading order in $1/N_s$ and 
	in the diagrams B -- N at the next-to-leading order. 
There is no mixing between 
	the gauge boson line with one-scalar-loops inserted (Fig.\ \ref{f2}a)
	and the scalar-loop chain due to $\lambda _{\rm r}|\phi _{{\rm r}j}|^4$
	interaction (Fig.\ \ref{f2}b),
	because the sub-diagram in Fig.\ \ref{f3}a vanishes.
In calculation, we adopt the minimal subtraction prescription
	in the dimensional regularization.
To avoid the absurdity of vanishing coupling constants,
	we should keep $\epsilon =(4-d)/2$ at a very small but non-vanishing value,
	where $d$ is the analytically continued number 
	of the spacetime dimensions.
We retain only the part leading in $\epsilon $,
	which amounts to retaining the terms of $O(\epsilon ^0)$
	because the compositeness condition implies $g_{\rm r}^2=O(\epsilon )$ 
	and each diagram has at most
	the order of magnitude of $g^{2n}/\epsilon ^n=O(\epsilon ^0)$. 

Cancellations between the sub-diagrams in Fig.\ \ref{f3}b
	extremely facilitate the calculations.
Because the one-scalar-loop (denoted by $\Pi _0^{\mu \nu }$) inserted 
	into a gauge-boson propagator (with momentum $q$) (Fig.\ \ref{f2}a)
	is divergenceless (i.e. $q_\mu \Pi _0^{\mu \nu }=0$),
	it suppresses the $\alpha _{\rm r}$-dependent part of the diagram.
After lengthy calculations similar to those indicated in Ref.\ \cite{cgb2}, 
	we obtain the contributions from the A -- H to $Z_3$
	as are shown in Table 1,
	where we also cited the corresponding results for
	the model with fermionic matters in the previous paper \cite{cgb2}.
The diagrams I and J in Fig.\ \ref{f1} contribute no leading divergences,
	because the scalar-loop chain behaves like $(-q^2)^{-l\epsilon }$,
	($q$ is the total momentum transfer through the chain,
	and $l$ is the number of loops in the chain.)
	and consequently the integration over $q$ 
	give rise to an extra factor of $\epsilon $ 
	in comparison with the leading contributions in $\epsilon $.
The diagram K is not leading in $\epsilon $ 
	because the sub-diagram Fig.\ \ref{f3}c converges,
	and the diagrams L -- N cancel out because 
	the sub-diagrams in Fig.\ \ref{f3}d cancel,

Next, we renormalize the sub-diagram divergences by
	subtracting the divergent counter parts of 
(i) each scalar loop inserted into the gauge-boson lines in D -- H, 
(ii) the scalar self-energy part in D,
(iii) the scalar-scalar-gauge-boson vertex parts in D$'$, E, G, and H, and
(iv) the three-gauge-boson vertex part in G and H. 
The contributions from the counter parts cancel out for 
	the diagrams with odd loops,
	and amount to minus twice the original terms for 
	those with even loops.
Collecting all the contributions in Table 1 and the counter terms,
	we finally obtain the compact expression 
\begin{eqnarray} 
Z_3 &=& 1 - { 1\over 6} N_s g_{\rm r}^2 I
         + {11\over 3} N_c g_{\rm r}^2 I
         - {\alpha _{\rm r}\over 2} N_c g_{\rm r}^2 I 
	(1 - { 1\over 6} N_s g_{\rm r}^2 I)
\cr &&
   + {3\over 2} N_c ( {6\over N_s} - g_{\rm r}^2 I)
      \ln(1 - { 1\over 6} N_s g_{\rm r}^2 I)
    +O({1\over N_s^3}),
\label{Ztot}
\end{eqnarray} 
where $I= {1}/{16\pi^2\epsilon}$.
It is remarkable that this is much similar in form
	to the corresponding result for the previous model 
	with fermionic matters (\cite{cgb2}), in spite that 
	the contributions from the corresponding diagram differ.
In fact (\ref{Ztot}) is what is obtained by replacing $N_f$ by $N_s/4$ 
	in the result Eq.\ (12) in \cite{cgb2}.
We note that $Z_3$ does not depend 
	on the scalar self-coupling constant $\lambda_{\rm r}$.

The compositeness condition $Z_3=0$ with expression (\ref{Ztot})
	can be solved for $g_{\rm r}^2$
	by iterating the leading-order solution into itself.
The solution is rather simple:
\begin{eqnarray} 
	g_{\rm r}^2 = {6\over N_sI}
	\left[  1+{22N_c\over N_s}+O({1\over N_s^2})\right] .
\label{g}
\end{eqnarray} 
The logarithmic term in (\ref{Ztot}) is suppressed in the solution (\ref{g})
	by the factor which vanishes in iteration of the leading solution.
It is interesting that the solution does not depend on the 
	gauge parameter $\alpha _{\rm r}$, 
	in spite of the fact that $Z_3$ does.
The $\alpha _{\rm r}$-dependent term in (\ref{Ztot}) 
	is also suppressed in the solution
	in the same way that the logarithmic term is.
The solution of the compositeness condition should be gauge-independent
	because it is a relation among physically observable quantities.

Because the above argument relies on $1/N_s$ expansion including iteration,
	the absolute value of the next-to-leading contribution
	should not exceed that of the leading one.
If we apply it to (\ref{g}), we obtain
\begin{eqnarray} 
	N_s>{22N_c}.
\label{NfNc}
\end{eqnarray} 
The allowed region of $N_s/N_c$ by (\ref{NfNc})
	is complementary to that for asymptotic freedom 
	in the gauge theory \cite{af},
	just like in the case of the previous model 
	with fermionic matters \cite{cgb2}.
When the gauge theory is asymptotically free,
	the next-to-leading contributions to the compositeness condition
	are so large that the gauge bosons cannot be composites 
	of the above type.
On the contrary, when the theory is asymptotically non-free, 
	the next-to-leading order contributions are suppressed,
	and the gauge bosons can be interpreted as composite.

Though the complementarity in the scalar matter system
	is very parallel to that in the fermionic system, 
	they are much different in several points.
Because the scalar mass is not protected by any symmetry
	unlike fermions',
	quadratic divergences arise in the scalar self-mass.
Though they are not explicit in the dimensional regularization,
	they should be interpreted to exist from the physical points of view.
Then they should be renormalized by unnatural fine tuning.
Another characteristic of the scalar matter system is 
	the existence of the self coupling constant $\lambda$.
At low energies in comparison with the cutoff scale,
	$\lambda$ does not appear in the compositeness condition,
	as is mentioned above. 
It could, however, affect our result in various way.
For example, it runs with the energy scale, 
	and may blow up at some scale, 
	invalidating the approximation at the order in $1/N_s$.
On the other hand, the gauge coupling constant $g$
	remains fixed even for large scale $\mu$ 
	(but below the cutoff $\Lambda\sim m_{\rm r}e^{1/\epsilon}$),
	because the renormalization group beta function
\begin{eqnarray} 
	\beta(g_{\rm r})\equiv\mu{\partial g_{\rm r}\over\partial\mu}
	=-\epsilon g_{\rm r}+{1\over6}N_sg_{\rm r}^3I-{11\over3}N_cg_{\rm r}^3I.
\end{eqnarray} 
	vanishes due to solution (\ref{g}) of the compositeness condition
	within the present approximation.
In general, however, $g$ would also blow up together with $\lambda$
	at the order where $Z_3$ depends on $\lambda$.

It is straightforward to extend this result to the case
	where both the fermionic and scalar matters exist.
We have only to replace $N_s$ in (\ref{Ztot}) and (\ref{g}) by $N_s+4N_f$,
	and consequently the complementality still exists
	because the same replacement rule holds 
	also for the asymptotic freedom.
It is tempting to extend it to supersymmetric models.
For this case, there are a few problems to be overcome.
First we should take into account the effects of the gaugeno, 
	the superpartner of the gauge boson.
It gives rise to further quantum corrections 
	through intermediate virtual states,
	and at the same time it should be a composite
	since it is a partner of the composite gauge boson.
Furthermore it has an additional interaction through 
	gaugeno-scalar-fermion vertex,
	which would give rise to complexities of diagrams
	contributing to $Z_3$,
	while asymptotic freedom seems to be free of the effects
	of these additional interactions.
These problems are now under investigation.

We would like to thank Professor H. Terazawa for useful discussions.
This work is supported by Grant-in-Aid for Scientific Research, 
	Japanese Ministry of Culture and Science.

\newpage

\noindent 
{\bf Table 1. \ Contributions to $Z_3$ from the diagrams A -- H} \\
The diagrams A -- H are defined 
in Fig.\ 1 in Ref.\ \cite{cgb2} for the model with spinorial matters and 
in Fig.\ 1 in this paper for the model with scalar matters.
$I=1/16\pi ^2\epsilon $.

\renewcommand{\arraystretch}{1.3}
\def\fcircle {\displaystyle}
\begin{center}
\begin{tabular}{l|l|l} 
%\multicolumn{3}{c}{ Table.1 \ \ \  } 
diagram&with spinorial matter&with scalar matter
\\ \hline%-------------------------------------------------
  A   
&$\fcircle -{2\over 3}N_fg_{\rm r}^2 I$
&$\fcircle -{1\over 6}N_sg_{\rm r}^2 I$
\\ \hline%-------------------------------------------------
B, C
&$\fcircle \left( {13\over 6}-{\alpha _{\rm r}\over 2}\right) N_cg_{\rm r}^2 I$
&$\fcircle \left( {13\over 6}-{\alpha _{\rm r}\over 2}\right) N_cg_{\rm r}^2 I$
\\ \hline%-------------------------------------------------
D, D$'$, E
&$\fcircle {1\over 3 }\alpha _{\rm r} N_c N_f g_{\rm r}^4 I^2$
&$\fcircle {3\over 2} N_c \sum _{l=1}^\infty \left( -{1\over 6}\right) ^l{N_s^l (g_{\rm r}^2 I)^{l+1}\over l(l+1)}$
\\&&
$\fcircle +{1\over 12}\alpha _{\rm r} N_c N_s g_{\rm r}^4 I^2$
\\ \hline%-------------------------------------------------
F, G, H
&$\fcircle {3\over 2} N_c \sum _{l=1}^\infty \left( -{2\over 3}\right) ^l{N_f^l (g_{\rm r}^2 I)^{l+1}\over l(l+1)}$
&$\fcircle -{1\over 6 }\alpha _{\rm r} N_c N_s g_{\rm r}^4 I^2$
\\&
$\fcircle -{2\over 3}\alpha _{\rm r} N_c N_f g_{\rm r}^4 I^2$&

\end{tabular}
\end{center}

\newpage



\begin{thebibliography}{99}
\bibitem{cg}     % H Bj ref AH
W.~Heisenberg,                  Rev.\ Mod.\ Phys.\ {\bf 29}, 269 (1957);
J.~D.~Bjorken,                  Ann.\ Phys.\ {\bf 24}, 174 (1963).
%%CITATION = APNYA,24,174;%% 
I.~Bialynicki-Birula,           Phys.\ Rev.\ {\bf 130}, 465 (1963);
D.~Luri\'e and A.~J.~Macfarlane,  Phys.\ Rev.\ {\bf 136} (1964) B816;
T.~Eguchi,  {Phys.\ Rev.} {\bf D14} (1976) 2755; {\bf D17} (1978) 611;
%%CITATION = PHRVA,D14,2755;%%
K.~Shizuya,  {Phys.\ Rev.} {\bf D21} (1980) 2327.
%%CITATION = PHRVA,D21,2327;%% 



\bibitem{cgw}     
T.~Eguchi and H.~Sugawara, {Phys.\ Rev.} {\bf D10} (1974) 4257;
%%CITATION = PHRVA,D10,4257;%% 
H.~Kleinert, Phys.\ Lett.\ {\bf 59B} (1975) 163;
%%CITATION = PHLTA,B59,163;%%
A.~Chakrabarti and B.~Hu, {Phys.\ Rev.} {\bf D13} (1976) 2347.
%%CITATION = PHRVA,D13,2347;%% 
K.~Akama and H.~Terazawa, INS-Rep-{\bf 257} (1976); 
H.~Terazawa, Y.~Chikashige and K.~Akama, {Phys.\ Rev.} {\bf D15} (1977) 480;
%%CITATION = PHRVA,D15,480;%%
T.~Saito and K.~Shigemoto, {Prog.\ Theor.\ Phys.} {\bf 57} (1977) 242;
%%CITATION = PTPKA,57,242;%% 
O.~W.~Greenberg and J.~ Sucher, Phys.\ Lett.\ {\bf 99B} (1981) 339; 
%%CITATION = PHLTA,B99,339;%% 
L.~F.~Abbott and E.~Farhi, Phys.\ Lett.\ {\bf 101B} (1981) 69; 
%%CITATION = PHLTA,B101,69;%% 
K.~Akama and T.~Hattori, Phys.\ Rev.\ {\bf D39} (1989) 1992;
{\bf D40} (1989) 3688;Int.\ J.\ Mod.\ Phys.\ {\bf A9} (1994) 3503.
%%CITATION = PHRVA,D39,1992;%%
%%CITATION = PHRVA,D40,3688;%%
%%CITATION = IMPAE,A9,3503;%%

\bibitem{indgra}
A.~D.~Sakharov,  Dokl.\ Akad.\ Nauk SSSR {\bf 177} (1967) 70
	[{Sov.\ Phys.\ Dokl.} {\bf 12} (1968) 1040];
%%CITATION = SPHDA,12,1040;%% 
K.~Akama, Y.~Chikashige and T.~Matsuki,
{Prog.\ Theor.\ Phys.} {\bf 59} (1978) 653;
%%CITATION = PTPKA,59,653;%%
K.~Akama, Y.~Chikashige, T.~Matsuki and H.~Terazawa,
	 {Prog.\ Theor.\ Phys.} {\bf 60} (1978) 868;
%%CITATION = PTPKA,60,868;%%
K.~Akama,  {Prog.\ Theor.\ Phys.} {\bf 60} (1978) 1900; 
Prog.\ Theor.\ Phys.\  {\bf 61} (1979), 687;
Phys.\ Rev.\ {\bf D24} (1981), 3073; 
%%CITATION = PTPKA,60,1900;%% 
%%CITATION = PTPKA,61,687;%%
%%CITATION = PHRVA,D24,3073;%%
S.~L.~Adler, Phys.\ Rev.\ Lett.\ {\bf 44} (1980) 1567;
%%CITATION = PRLTA,44,1567;%%
A.~Zee, Phys.\ Rev.\ {\bf D23} (1981) 858;
%%CITATION = PHRVA,D23,858;%%
D.~Amati and G.~Veneziano, Phys.\ Lett.\ {\bf 105B} (1981) 358;
			Nucl.\ Phys.\ {\bf B240} (1982) 451;
%%CITATION = PHLTA,B105,358;%%
%%CITATION = NUPHA,B204,451;%%
K.~Akama and I.~Oda,	Phys.\ Lett.\ B {\bf 259} (1991) 431;
Nucl.\ Phys.\ B {\bf 397} (1993) 727.
%%CITATION = PHLTA,B259,431;%%
%%CITATION = NUPHA,B397,727;%%


\bibitem{hidden}
M.~Bando, T.~Kugo, S.~Uehara, K.~Yamawaki and T.~Yanagida,
	Phys.\ Rev.\ Lett.\ {\bf 54} (1985) 1215. 
%%CITATION = PRLTA,54,1215;%% 
For vector meson as gauge boson, 
see J.~J.~Sakurai, {\it Currents and Mesons}, 
(Univ. Chicago Press, Chicago, 1969).
For the hidden local symmetry,
E.~Cremmer and B.~Julia, Phys.\ Lett.\ {\bf 80B} (1978) 48; 
	Nucl.\ Phys.\ {\bf B159} (1979) 141.
%%CITATION = PHLTA,B80,48;%% 
%%CITATION = NUPHA,B159,141;%%


\bibitem{emb}
K.~Akama, Lect. Notes in Phys.\ {\bf 176} (1983) 267; 
{Prog.\ Theor.\ Phys.} {\bf 78} (1987) 184; 
{\bf 79} (1988) 1299; 
{\bf 80} (1988) 935; 
hep-th/0307240 (2003);
%%CITATION = HEP-TH 0001113;%%
%%CITATION = PTPKA,78,184;%%
%%CITATION = PTPKA,79,1299;%%
%%CITATION = PTPKA,80,935;%%
%%CITATION = HEP-TH 0307240;%%
K.~Kikkawa, Phys.\ Lett.\ {\bf B297} (1992) 89; 
%%CITATION = PHLTA,B297,89;%% 
G.~Dvali, G.\ Gabadadze, and M.\ Porrati, Phys.\ Lett.\ {\bf B485} (2000) 208; 
%%CITATION = HEP-TH 0005016;%%
K.~Akama and T.~Hattori, Mod.\ Phys.\ Lett.\ A {\bf 15} (2000) 2017.
%%CITATION = HEP-TH 0008133;%%


\bibitem{qm}
N.~P.~Landsman and N.~Linden, Nucl.\ Phys.\ {\bf B365} (1991) 121; 
%%CITATION = NUPHA,B365,121;%%
Y.~Ohnuki and S.~Kitakado, J.\ Math.\ Phys.\ {\bf 34} (1993) 2827; 
%%CITATION = JMAPA,34,2827;%% 
D.~McMullan and I.~Tsutsui, Phys.\ Lett.\ {\bf B320} (1994) 287; 
		Ann.\ Phys.\ {\bf 237} (1995) 269; 
%%CITATION = HEP-TH 9310185;%%
%%CITATION = HEP-TH 9308027;%%
S.~Tanimura and I.~Tsutsui, Mod.\ Phys.\ Lett.\ {\bf A10} (1995) 2607. 
%%CITATION = HEP-TH 9508165;%% 


\bibitem{geo}
C.~A.~Mead and D.~G.~Truhlar, J.\ Chem.\ Phys.\ {\bf 70(05)} (1979) 2284;
M.~V.~Berry, Proc.\ R.\ Soc.\ London Ser.\ {\bf A392} (1984) 45;
%%CITATION = PRSLA,A392,45;%%
A.~Shapere and F.~Wilczek, ``{\it Geometric Phases in Physics"}
	(World Scientific, 1989).


\bibitem{cc}       
B.~Jouvet, 		{Nuovo Cim.} {\bf 5} (1956) 1133; 
M.~T.~Vaughn, R.~Aaron and R.~D.~Amado, 
			{Phys.\ Rev.} {\bf 124}  (1961) 1258;
A.~Salam, 		{Nuovo Cim.} {\bf 25} (1962) 224; 
%%CITATION = NUCIA,25,224;%%
S.~Weinberg, 		{Phys.\ Rev.} {\bf 130} (1963) 776;
%%CITATION = ASJOA,168,175;%%
Luri\'e and Macfarlane,  Ref.\ \cite{cg};
T.~Eguchi, Ref.\ \cite{cg};
H.~Kleinert, in {\it Understanding the fundamental
constituents of matter, proceedings, 1976 Erice Summer School}, 
ed. A. Zichichci (Plenum Publishing Corporation, 1978), 289;
K.~Shizuya, Ref.\ \cite{cg}.

\bibitem{ccA}
K.~Akama,  		{ Phys.\ Rev.\ Lett.} {\bf 76} (1996) 184;
Nucl.\ Phys.\ A {\bf 629} (1998) 37C.
%%CITATION = HEP-PH 9511301;%%
%%CITATION = NUCL-TH 9709001;%%

\bibitem{cgb1}
K.~Akama and T.~Hattori, { Phys.\ Lett.} {\bf B392} (1997) 383. 
%%CITATION = HEP-PH 9607331;%%

\bibitem{cgb2}
K.~Akama and T.~Hattori,  { Phys.\ Lett.} {\bf B445} (1998) 106.
%%CITATION = PHLTA,B445,106;%%

\bibitem{af}
D.~J.~Gross and F.~Wilczek, { Phys.\ Rev.\ Lett.} {\bf 30} (1973) 1343;
%%CITATION = PRLTA,30,1343;%%
H.~D.~Politzer, { Phys.\ Rev.\ Lett.} {\bf 30} (1973) 1346.
%%CITATION = PRLTA,30,1346;%% 
\end{thebibliography}
\end{document}